\documentclass[epj,referee]{svjour}
\usepackage{graphics}
\usepackage{ulem}
\usepackage{color}

\begin{document}
\title{Trapped electron coupled to superconducting devices}
\author{P. Bushev\inst{1} \and D. Bothner\inst{2} \and J. Nagel\inst{2} \and M. Kemmler\inst{2} \and K. B. Konovalenko\inst{2}
\and A.L\"orincz\inst{3} \and K.Ilin\inst{3} \and M. Siegel\inst{3}
\and D. Koelle\inst{2} \and R. Kleiner\inst{2} \and F. Schmidt-Kaler\inst{4} 
}                     
\institute{Physikalisches Institut, Karlsruher Institut f\"ur Technologie, Wolfgang-Gaede-Str. 1, D-76131 Karlsruhe
Germany \and
Physikalisches Institut and Center for Collective Quantum Phenomena, Universit\"at T\"ubingen, Auf der Morgenstelle 14, 72076 T\"ubingen, Germany \and
Institut f\"ur Mikro- und Nanoelektronische Systeme, Karlsruher Institut f\"ur Technologie,  Hertzstrasse 16, 76187 Karlsruhe, Germany \and
Institut f\"ur Physik, QUANTUM, Universit\"at Mainz, Staudingerweg 7, 55128 Mainz, Germany}

\date{Received: date / Revised version: date}
%
\abstract{We propose to couple a trapped single electron to superconducting structures located at a variable distance from the electron. The electron is captured in a cryogenic Penning trap using electric fields and a static magnetic field in the Tesla range. Measurements on the electron will allow investigating the properties of the superconductor such as vortex structure, damping and decoherence. We propose to couple a superconducting microwave resonator to the electron in order to realize a circuit QED-like experiment, as well as to couple superconducting Josephson junctions or superconducting quantum interferometers (SQUIDs) to the electron. The electron may also be coupled to a vortex which is situated in a double well potential, realized by nearby pinning centers in the superconductor, acting as a quantum mechanical two level system that can be controlled by a transport current tilting the double well potential. When the vortex is trapped in the interferometer arms of a SQUID, this would allow its detection both by the SQUID and by the electron.} 
\maketitle
\section{Introduction}
\label{intro}
Single atomic quantum systems, photons, ions or solid state systems such as Josephson junctions, SQUIDs, color centers or micro mechanical oscillators, are mastered today almost in perfection. As their quantum properties are illuminated in detail, we face the new challenge to combine either many identical copies of such elementary systems or to combine different ones to investigate the enriched properties of such hybrid quantum systems.

Examples of hybrid quantum systems are single ions coupled by an electric conductor \cite{Soerensen04,Tian04,Haeffner09b}, Rydberg atoms interacting with a superconducting microwave resonator \cite{Gleyzes07} and cold atoms above superconducting traps \cite{Mukai07,Nirrengarten06,Kasch10,Cano08,Mueller10} or carbon nanotubes \cite{Gruener09}. Coplanar superconducting resonators with high quality factors $Q$ have been successfully coupled to a Cooper pair box \cite{Wallraff04}; it is intended to exploit a similar technique to coherently couple trapped molecules \cite{Andre06} and Bose-Einstein condensates \cite{Verdu09} to superconducting cavities. Very recently strong cooperative magnetic coupling has been demonstrated with spin ensembles implemented with nitrogen-vacancy(NV) centers in diamond \cite{Kubo10}, and with both ruby(Cr$^{3+}$ ions) and NV defects \cite{Shuster10} coupled to the magnetic field of a single photon in a superconducting resonator. Other examples of hybrid quantum systems "naturally" appear in superconducting qubits and are usually associated with two-level microscopic defect states coupled to the electric field inside the Josephson junctions \cite{Martinis08,Lupascu09,Bushev10}. Approaching a single electron to a low temperature surface and measuring the decoherence properties of such systems would allow to complement nicely investigations of motional heating in planar ion traps, as performed with closeby normal conducting or superconducting surfaces at a temperature of 4.2\,K temperature \cite{ANTOLI2009,LABAZ2008}.

The integration of superconducting devices and traps for single electrons will combine a fundamental solid state quantum system with a controlled quantum optical system. Both systems are operated in a very low noise environment and at low temperatures provided by a $^3$He/$^4$He dilution refrigerator. Such a hybrid system may improve the understanding of the solid state component by measurements taken from the electron. On the technological side, common requirements can be met, such as the need for low temperatures $T$ of a few 10\,mK and the demand for high $Q$ values, which can be achieved, when the materials are cooled down to low temperatures. The fabrication of devices in the form of two-dimensional micro structures, integrated in a chip design, is required. Only then the trapped electron can be precisely moved to the close vicinity of the solid state system. Furthermore the energy scales of both systems are comparable, and for their mutual coupling tuning knobs are at hand to fine-adjust frequencies and coupling strengths. Still, many problems remain to be solved, and we address some of them in this proposal. They regard the single electron trapping and detection in planar micro structured devices, the fabrication of superconducting devices which can be operated in Tesla magnetic fields, as well as the experimental integration complexity, especially for a low temperature experiment, and the coupling schemes.

The paper starts with a brief introduction of the specialized Penning electron trap (Sect.\ref{trap}) and continues with some remarks on general design and fabrication requirements for the superconducting elements (Sect.\ref{SC}). Then we sketch a few ideas for interacting electron-superconductor hybrid systems and discuss the mutual coupling strengths (Sect.\ref{couple}).
Section 5 contains our conclusions.

\section{Adapted single electron trap}\label{trap}
\begin{figure}[tb]
\begin{center}
\resizebox{0.9\columnwidth}{!}{%
 \includegraphics{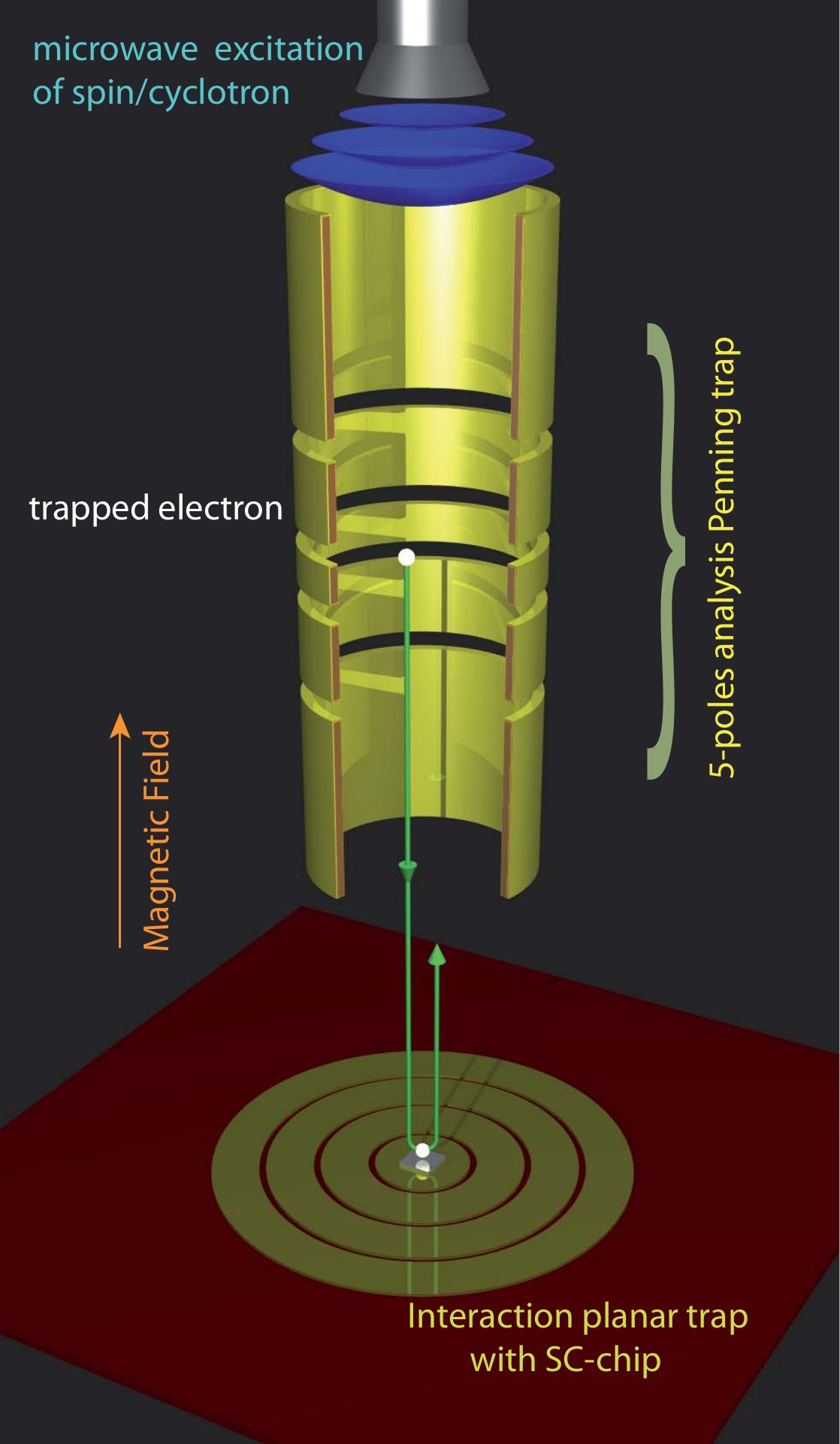} }
\end{center}
\caption{(color online) The sketch of the experimental setup to measure the interaction between an electron and a SC-surface or a SC-chip denoted as a small square inside the central ring electrode of the planar trap. The magnetic field is directed upwards. The LC-tank circuit for cooling down axial motion and DC-filters for voltages on trap electrodes are omitted for simplicity.
}
\label{fig:setup}
\end{figure}

While three-dimensional single electron traps \cite{GabiRMP86} have been operated for many decades, e.g. leading to precision experiments \cite{Quint96,Werth2000,Werth2002,Gabrielse08} for the determination of the g-factor or the mass of a single electron and for accurate QED tests, the interest in planar Penning traps was triggered by proposals to use the electron spin for quantum information processing \cite{Tombesi2003}, and corresponding research activities have only started recently \cite{Unsere2008,GabiPRA2010,GOLD2010}. The advantage of planar Penning traps, i.e. short distance to trap electrodes, however, might also regard other applications, as micro fabrication technologies, which pave a route for a better scalability of devices, combined with unmatched fabrication precision and surface cleanness. Until today, only small clouds of electrons have been detected in cryogenic micro planar Penning traps \cite{Unsere2008}, and single electron detection has not been demonstrated so far in such micro-traps. To benefit from \textit{both}, the established technique of single electron trapping and detection in a three-dimensional Penning trap and a micro-structured planar device, we propose a combination of both in one setup, see Fig. \ref{fig:setup}.

The three-dimensional trap consists of a series of concentric ring electrodes which are supplied by well chosen DC voltages to form a quadrupole potential above the surface. A homogeneous magnetic field $B_0$ is applied perpendicular to the plane. The electron is confined and constitutes a harmonic oscillator with frequencies
\begin{eqnarray}
  \omega_{+} &=& \omega_c + \omega_z^2/2 \omega_c~~, \\
  \omega_{-} &=& \omega_z^2/2 \omega_c~~, \\
   \omega_z &=& \sqrt{2eU/md^2}~~,
\end{eqnarray}

where $\omega_c = e B_0 / m c$ is the free electron cyclotron frequency, $\omega_+$ is the reduced cyclotron frequency, $\omega_-$ the magnetron frequency, and $\omega_z$ the axial frequency. Characteristic numbers for these frequencies in a trap of size $r_0\sim $ 1\,mm, a voltage $U \sim $~5\,V and $B_0 \sim$ 1\,T are $\omega_+/(2\pi) \sim $~30\,GHz, $\omega_-/(2\pi) \sim $~150\,kHz and $\omega_z/(2\pi) \sim$~100\,MHz. Once a single electron is trapped in such a potential well, it equilibrates with the thermal environment at a few 10\,mK temperature in the cyclotron degree of freedom, such that the oscillation is described by a quantum oscillator, almost perfectly prepared in its ground state.

The harmonicity of the axial potential is of upmost importance for the single electron detection and can be adjusted by the proper size and shape of the ring electrodes and their control voltages. An analytic calculation for planar Penning traps is found in \cite{GOLD2010}, specializing the design rules which have been introduced in \cite{GabiRMP86} and used for three-dimensional traps \cite{Walz}. A numerical code would take into account non-symmetric contributions, electrode gaps and other imperfections \cite{Hellwig,rmpKilian}. When the electron is detected with a resonant circuit and a pre-amplifier anchored at $T = $ 0.1\,K, the axial electron temperature of typically about 5\,K results from its coupling to the thermal noise of the amplifier, which noise temperature can hardly be lowered into the sub-Kelvin range, see Refs. \cite{Ukhanskij2004,Pospizalski}. Thus, the electron wavefunction experiences some range of the potential, sensing also its anharmonicities. Only if such anharmonic terms are corrected for, the axial resonance remains narrow, such that the single electron resonance is detectable.

As the demands on detection of a single electron and preparation of its quantum state are quite different from the conditions, when the electron is brought into interaction with nearby superconducting surfaces or surface devices, we propose to shuttle the electron between two zones in a multi-segmented  Penning trap. In one instant, the electron quantum state is detected, later it is brought in close vicinity to the solid state devices. This technique of shuttling has been applied very successfully for precision measurements of the g-factor of an electron bound in hydrogenlike carbon in nested Penning traps \cite{Werth2000}; Fig. \ref{fig:setup} sketches this method. In contrast to the other interesting proposals for manipulating spin of electrons on liquid helium surface \cite{SchusterElectron2010}, the 3D-Penning trap offers properties to store the trapped particles for the very long time (months) and a low-decoherence environment.

The shuttling experimental sequence can be performed in the following way: the electron is initially captured and cooled down in the "analysis" three-dimensional trap. A strong, cyclotron $\pi/2$ pulse is applied together with its anomalous drive and creates a superposition $|\uparrow\rangle+i|\downarrow\rangle$ of electronic spin states. By applying a proper change of the trap potentials, the trapped electron is shifted towards the interaction trap, where it gets trapped and cooled again. After an interaction time $\tau$, electron is moved back into the "analysis" trap and the spin state is read out with an additional $\pi/2$ pulse, followed by a measurement of its axial frequency \cite{Odom2006}. By repeating this sequence for varying interaction time, detuning and distance to the surface of the interaction planar trap, one can perform quantum decoherence microscopy \cite{Jared09} and obtain the full information about spin-surface interaction.

\section{Superconducting structures}\label{SC}
Combining single electrons with superconducting structures requires the application of static magnetic fields $B$ in the Tesla range, which is much higher than the fields, in which typical active or passive superconducting devices are operated (earth magnetic field or below). One thus faces restrictions on the superconducting materials as well as on the design of the various devices. Josephson junctions, SQUIDs or resonators are most reliably made from either Nb or, in the context of superconducting qubits operated at Millikelvin temperatures, from Al.  At Tesla fields, Al is not superconducting.  If the thin film devices of interest can be oriented such that the applied magnetic field is as parallel as possible to the thin film surface, Nb can be considered for fields up to $B=$ 2\,T or so. For all other purposes one should either consider high temperature superconductors such as YBa$_2$Cu$_3$O$_7$ or use metallic compounds such as NbN that are able to sustain several Tesla. In any case, the superconductor to be used is type II and typically will be in the vortex state at the operating conditions. In a magnetic field of $B=$ 0.1\,T oriented perpendicular to the substrate plane, the intervortex distance is 150\,nm, decreasing to 45\,nm at $B=$ 1\,T. Motion of Abrikosov vortices leads to dissipation and magnetic field noise and will substantially degrade the device performance. To circumvent this problem, one either has to keep out vortices from the superconducting structures by making them small enough or one has to pin the vortices by, e. g., incorporating properly dimensioned artificial pinning sites into the structures. Below, examples for both strategies will be given.

\section{Coupling schemes}\label{couple}

\begin{figure}[tb]
\begin{center}
\resizebox{1.0\columnwidth}{!}{%
 \includegraphics{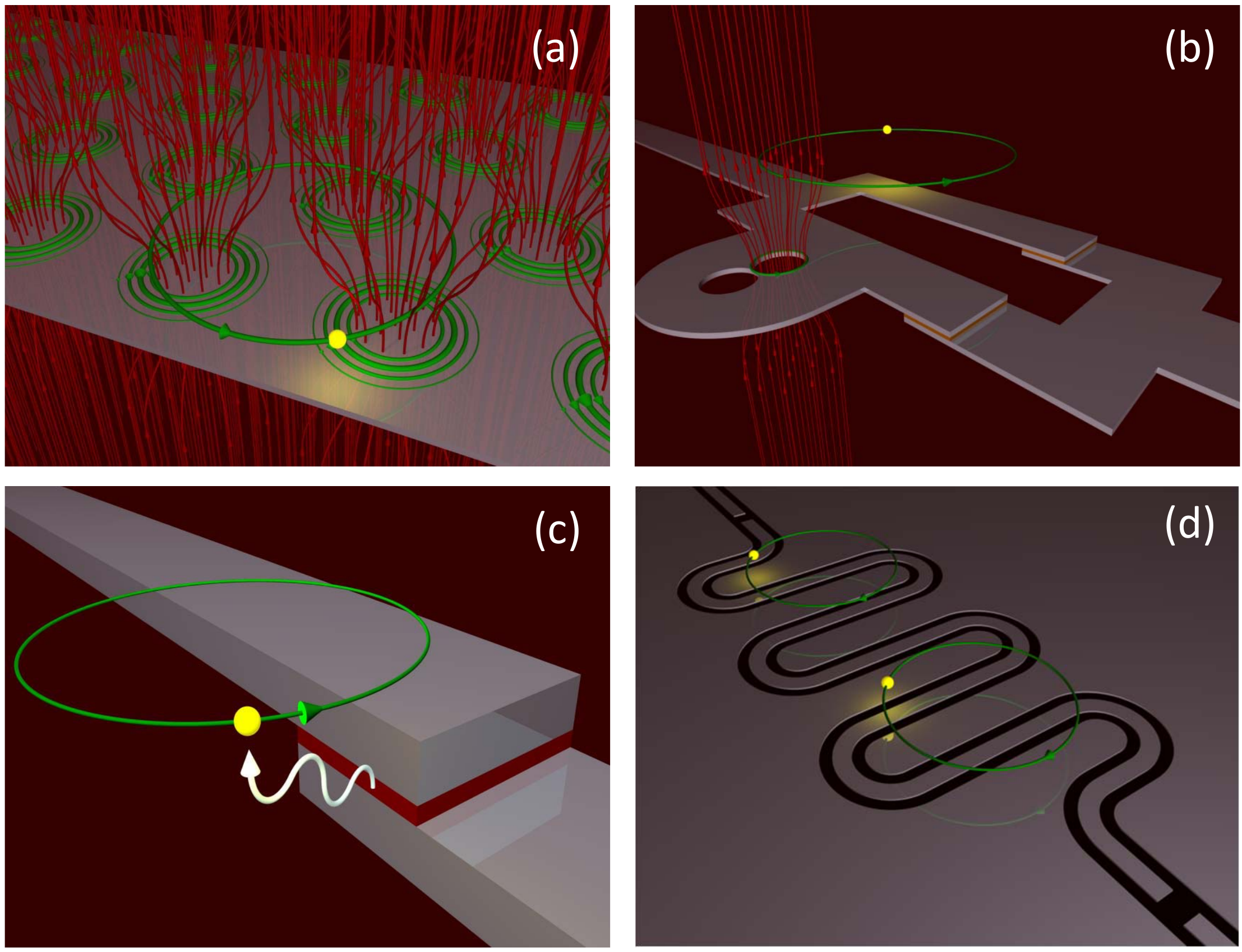} }
\end{center}
\caption{(color online) Illustrations of schemes to couple an electron (a) to vortices penetrating a superconducting film, (b) to a single vortex embedded in a SQUID, (c) to a Josephson junction and (d) to a superconducting resonator.
}
\label{fig:coupling}
\end{figure}


There are several ways to couple an electron to a superconducting structure. The motion of the trapped electron will be affected by magnetic fields created by the superconductor. In addition, when superconducting resonators or active devices such as Josephson junctions or SQUIDs are considered, coupling to electric fields is possible as well. Fig.\ref{fig:coupling} illustrates some coupling schemes schematically. In Fig.\ref{fig:coupling} (a) the electron orbits above a superconducting thin film which is penetrated by a lattice of Abrikosov vortices. In Fig.\ref{fig:coupling} (b) a SQUID is shown which has trapped a single Abrikosov vortex. The vortex affects the electron orbit; simultaneously, its dynamics can be monitored by the SQUID itself. Fig.\ref{fig:coupling} (c) illustrates a Josephson junction biased in its resistive state at a voltage $V$. According to the Josephson relation it radiates at a frequency $f=V/\Phi_0$, where $\Phi_0 \approx 2.07\cdot$10$^{-15}$\,Wb is the magnetic flux quantum. The electron can couple to the electromagnetic field generated by the oscillating Josephson currents. Finally, Fig.\ref{fig:coupling} (d) illustrates two electrons orbiting above a superconducting coplanar waveguide resonator.

\subsection{Magnetic coupling to Abrikosov vortices}
Let us first consider magnetic fields generated by the superconductor. They couple to the electron spin and also affect the electron orbit. Apart from having the possibility to use (super)current carrying structures (current biased lines or persistent currents flowing in a superconducting loop)
one can consider magnetic fields generated by Abrikosov vortices penetrating the superconductor.
This allows one to use the electron as a sensor to read out vortex properties and, ultimately, could lead to coupled vortex/electron states.

In an unpatterned film the electron will orbit above a more or less regular lattice of Abrikosov vortices, causing a periodic field modulation above the superconducting surface. The field generated by one vortex is on the order of $\Phi_0/\lambda_L^2$, where $\lambda_L$ (typically 100--200\,nm, depending on the superconductor) is the London penetration depth. For intervortex distances well above $\lambda_L$ one thus has a field modulation in the range of 100\,mT at the superconducting surface, which however decreases almost exponentially with growing distance from the surface. Also, for fields (perpendicular to the substrate) higher than 0.1\,T the vortices begin to overlap strongly and the field modulation further decreases.

In principle, the trapped electron can sense magnetic field variations at a level of few parts in 10$^{-13}$ \cite{Gabrielse08}, which would correspond to the sub-pico Tesla regime. Nevertheless, initially one should aim to reach micro- or nano-Tesla accuracy which is already enough to reveal the interaction with a superconducting surface.

\begin{center}
\begin{figure}[tb]
\resizebox{1\columnwidth}{!}{%
 \includegraphics{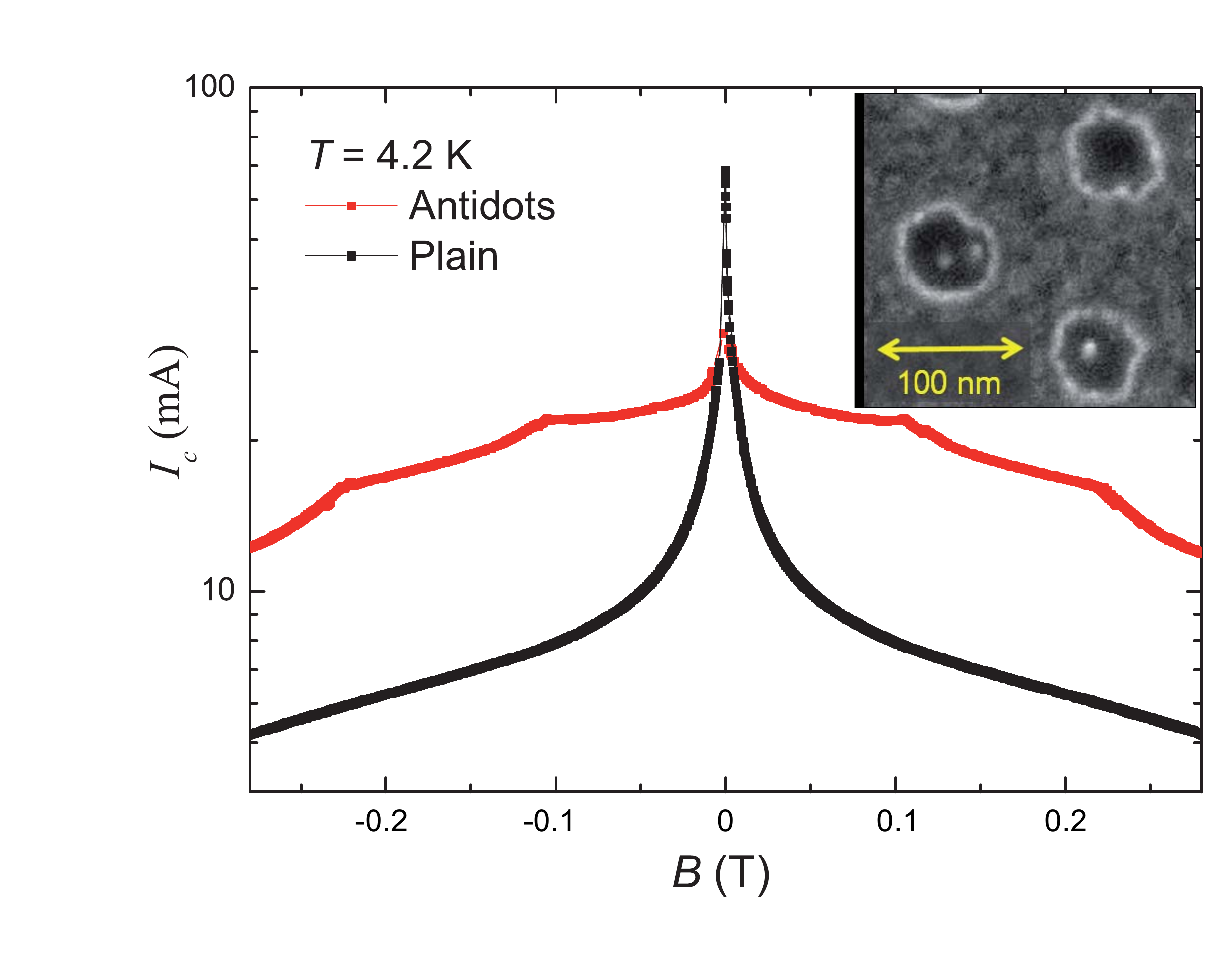} }
\caption{(color online) Critical current at 4.2\,K vs. magnetic field of a 40 x 70 $\mu$m$^2$ large bridge structure patterned into Nb thin films. The graph compares an unperforated (plain) film with a film containing about $10^5$ circular antidots (diameter 60\,nm) arranged in a triangular lattice (150\,nm lattice constant). The inset is a SEM image showing three antidots.
}
\label{fig:Antidots}
\end{figure}
\end{center}

If a certain configuration (ranging from triangular or square to quasiperiodic \cite{Kemmler06}) of vortices in the film plane is desired, one may pattern the film, e. g. by introducing nanoholes (antidots) hosting the vortices or by patterning arrays of superconducting rings. This will increase the field modulation at a given distance to the surface, although not by orders of magnitude. More importantly, the maximum supercurrent that can be carried by a perforated structure can be enhanced significantly.
Arrays of micrometer sized antidots have been studied intensively in the context of vortex pinning or vortex guidance \cite{Baert95,Woerdenweber04}. Recently, also hole sizes around 20\,nm have been reported in self-assembled systems \cite{Eisenmenger07}. One can also use arrays of vertically grown carbon nanotubes as columnar pinning sites with a diameter around 60\,nm \cite{Haeffner09}.

Figure \ref{fig:Antidots} shows data for a superconducting structure that contains antidots patterned into a 20\,nm thick Nb film by using a combination of e-beam lithography and reactive ion etching. The antidot diameter is 60\,nm, the lattice constant is 150\,nm.
The figure shows the magnetic field dependence of the maximum supercurrent $I_c(B)$ of the perforated film in comparison to a plain film, as measured at 4.2\,K on a bridge structure patterned into the film. While the critical current of the unperforated film decreases monotonously with increasing field, the perforated film shows clear ``matching'' peaks when the vortex lattice becomes commensurate with the antidot lattice. For example, at a field of 0.1\,T, $I_c$ is increased by a factor of 2.8 compared to the plain film, demonstrating the usefulness of antidot lattices.

Rather than coupling to a large ensemble of vortices it is perhaps more interesting to achieve coupling to a small number of vortices, ideally a single one.
Then, the electron orbit partially experiences the field generated by this vortex, causing slight changes in the characteristic frequencies of the electron motion. If the sensitivity to sense magnetic field variations is at the level of nano-Tesla, it will be possible to couple a single vortex to a single electron at the distance between them of a few hundreds of micrometers.

One way to achieve this, is to pattern a single superconducting island (some 10 nm in diameter) hosting only one vortex at a given perpendicular field in the Tesla range. Much larger superconducting structures can be used if the substrate plane is mounted almost perpendicular to the applied field such that the flux penetrating the structure is of order $\Phi_0$.  In this configuration, if the structure has a diameter much larger than $\lambda_L$, due to the screening currents in the superconductor, the magnetic field generated by the vortex resembles that of a monopole. Along the vortex axis it decays inversely proportional to the distance above the film and can be sensed over relatively long distances.

\subsection{Coupling to Josephson junctions and SQUIDS}

\begin{figure}[tb]
\begin{center}
\resizebox{1.0\columnwidth}{!}{%
 \includegraphics{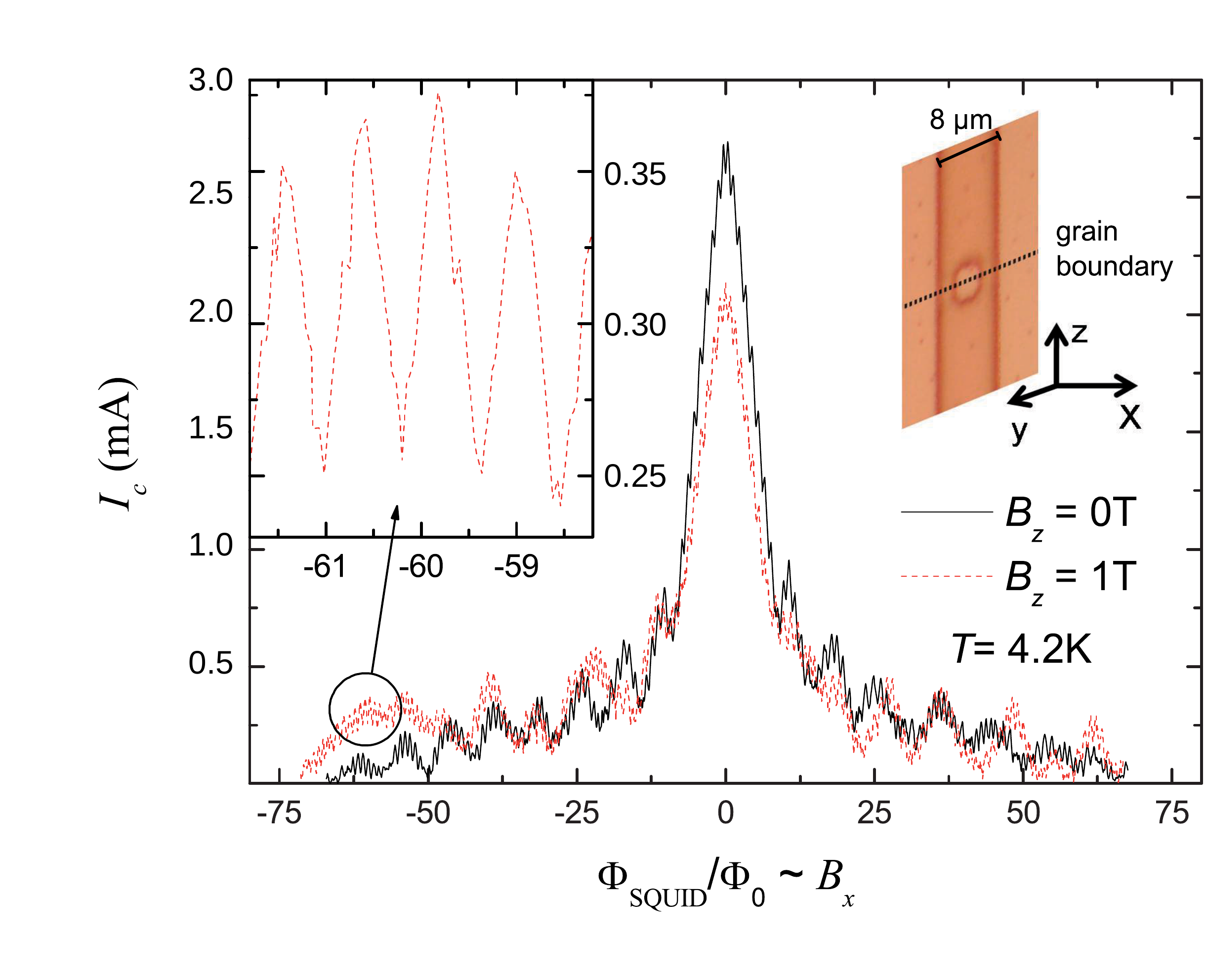} }
\end{center}
\caption{(color online) Modulation of the critical current of a YBa$_2$Cu$_3$O$_7$ SQUID, patterned on a SrTiO$_3$ bicrystal substrate, as a function of a magnetic field applied perpendicular to the substrate. In addition either no field (red line) or a 1\,T field (blue line) was applied parallel to the substrate (along $z$ axis, c. f. upper right inset). Upper left inset is an enlargement of the 1 T background field curve for fluxes between -61 $\Phi_0$ and -59 $\Phi_0$ penetrating the SQUID loop. Upper right inset shows an optical image of the device. Measurement temperature is $T=$ 4.2\,K.
}
\label{fig:YBCO_SQUID}
\end{figure}

A SQUID will most likely not be sensitive enough to detect a single electron via the magnetic field created by its spin or its orbital motion (which can be viewed as a circulating current coupling flux to the SQUID). However, still both SQUIDs and single junctions, will be useful elements to couple ac electromagnetic fields, generated by the ac Josephson currents at GHz frequencies, to the electron. SQUIDs can also be used as detectors for vortices in the superconducting structure. Taken to the extreme, one may consider coupling of the electron to a single Abrikosov vortex which is positioned in one of the arms of the SQUID loop.
The need to avoid junction degradation by the Tesla background field puts restrictions to the junction type. First, one should apply the field parallel to the substrate plane to avoid uncontrolled entry of vortices. For many standard junctions the superconducting and barrier layers are oriented parallel to the substrate plane. The critical current $I_c$ shows a Fraunhofer-type modulation as a function of a magnetic field applied parallel to the barrier layer, $I_c(B) = I_c(0)|\sin(x)/x|$, with $x=\pi \Phi_J/\Phi_0$. $\Phi_J$ denotes the magnetic flux penetrating the junction. The junction becomes in essence unusable when $\Phi_J$ exceeds several $\Phi_0$. Thus, the junctions either have to be extremely small or one chooses junctions having a barrier oriented perpendicular to the substrate. In the latter geometry the $I_c$ degradation is minimized if the field is oriented along the direction of current flow across the junction. Grain boundary junctions in high temperature superconductors are usable for this purpose.
Nb based micro-SQUIDs with a flux resolution of about 100 $\mu \Phi_0/$Hz$^{1/2}$ have been operated in background fields of up to 2 Tesla, oriented parallel to the substrate plane \cite{Wernsdoerfer00}. The Josephson junctions were realized by (hysteretic) weak links, presumably leading to the high values of flux noise (a flux resolution well below 1 $\mu \Phi_0/$Hz$^{1/2}$ is possible for the best SQUIDs \cite{Hao08}). Figure \ref{fig:YBCO_SQUID} shows data for a SQUID made of the high temperature superconductor YBa$_2$Cu$_3$O$_7$. The SQUID, which was not optimized in terms of size and inductance, was fabricated on a SrTiO$_3$ bicrystal substrate and contained two grain boundary Josephson junctions. The main graph shows the modulation of the critical current of the device as a function of a magnetic field $B_x$ applied perpendicular to the substrate. The large scale Fraunhofer-like variations in $I_c(B_x)$ are caused by the junctions; the high frequency oscillations are the actual SQUID modulations. The blue curve, taken in a 1\,T background field (oriented parallel to the substrate and in the direction  of current flow across the junction), does not differ much from the zero background field curve (red line), showing that these type of structures are usable in high background fields, which - for a suitably designed YBa$_2$Cu$_3$O$_7$ SQUID - may be as high as several Tesla.

In Fig.\ref{fig:coupling} (b) we have illustrated an example where two nearby antidots are embedded in a SQUID structure, allowing to capture a vortex in a controlled way. The two antidots create a double well potential for the vortex. It can be tilted, e. g., by passing a current, allowing to manipulate the vortex position, in turn causing back action on the electron. The SQUID is able to detect the motion of this vortex (indeed, in the thermal regime the corresponding random telegraph signals induced by the vortex often limit the SQUID performance), providing the readout of such a coupled vortex/electron system.

\subsection{Cavity QED-coupling}

\begin{figure}[tb]
\begin{center}
\resizebox{1.0\columnwidth}{!}{%
 \includegraphics{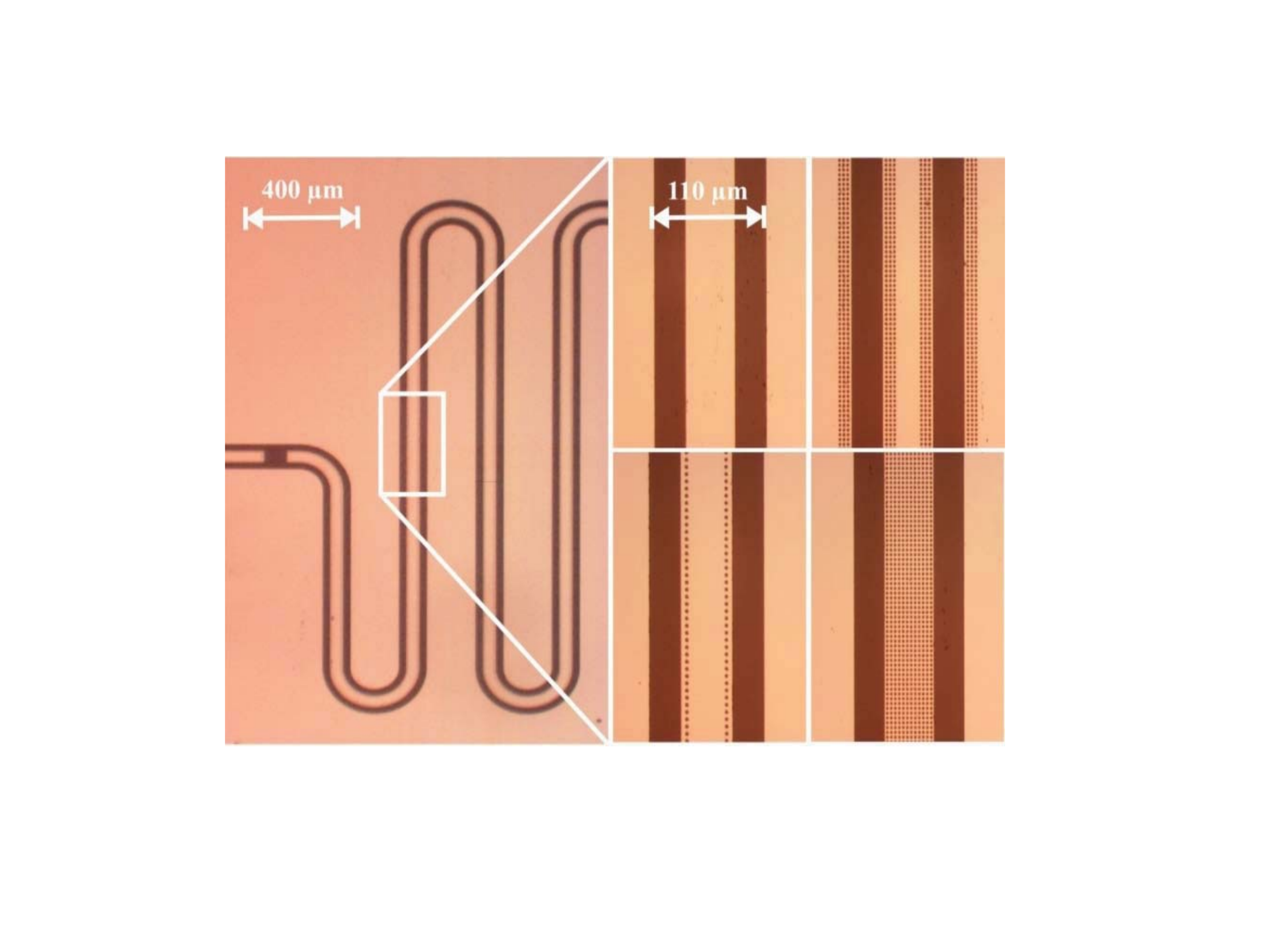} }

\end{center}
\caption{(color online) A Nb $\lambda$/2 thin film resonator perforated with antidots. Left: Full resonator, right: Zoom (made for 4 different samples) showing a plain structure and three different arrangements of antidots.
}
\label{fig:resonators}
\end{figure}
To couple axial or cyclotron degrees of freedom of an electron to a high-$Q$ resonator, like it has been already demonstrated for coupling of a Cooper-Pair-Box to a coplanar microwave resonator \cite{Wallraff04}, one needs to develop a suitable micro-Penning trap. This trap will have electrodes through-metalized on sapphire sheets, which makes its structure similar to a sandwich, thus limiting the stray capacity to a sub-pico Farad value. For efficient coupling of axial oscillations of an electron to the microwave resonator modes, the resonator must be implemented on one of the trap layers and be connected to the end cap of the penning trap. The loading of electrons into the trap can be performed with the help of an UV Light-Emission-Diode \cite{LED04}. The diameter of the trap is going to be in the range of 200-300\,$\mu$m, yielding the axial frequency in the range of 1-3\,GHz. The coupling strength of the axial degree of freedom to the electric field of the resonator is estimated to be on the order of 10-100\,kHz and can be measured in the transmission signal. To couple the electronic cyclotron rotation, the ring electrode or one half of the ring electrode will be connected to the microwave resonator, though the coupling strength in that case can hardly exceed a few 100\,Hz.

Resonator $Q$ factors well above $10^5$ can be achieved with superconductors at millikelvin temperatures without major problems at zero magnetic field. At high fields, for unperforated structures, there will be dissipation caused by vortices moving back and forth, due to Lorentz forces created by the supercurrents oscillating in the resonator. Thus, the technological task is to minimize this dissipation mechanism, both by introducing antidots and, if possible, by minimizing the field component perpendicular to the substrate plane. The situation is similar as for Josephson junctions and SQUIDs, although now dissipation has to be minimized for GHz frequencies. Further, resonator structures for frequencies of 3 GHz or lower are quite large, and the superconducting area is on the order of mm$^2$ if not cm$^2$. One thus faces the problem to either introduce a huge number of antidots or to pin only those vortices that create the strongest dissipation. Since the ac currents essentially flow at the edges of the superconducting structures, on the scale of $\lambda_L^2/d$ ($d$ is the film thickness), a few rows of antidots located near the edges may be sufficient.
The left graph in Fig.\ref{fig:resonators} shows the layout of a 3.3\,GHz Nb $\lambda$/2 coplanar waveguide resonator. At 4.2\,K, at zero magnetic field these resonators, when unperforated, reach $Q$ factors of some  10$^4$ \cite{Hammer07}. However, in a perpendicular field of 4\,mT $Q$ is already reduced to below 2000. The right hand graphs of Fig.\ref{fig:resonators} illustrate three different arrangements of antidots (one row along the two edges of the center conductor; three rows along the center and outer conductor; center conductor completely perforated). The antidot diameter used for these preliminary measurements was about 2\,$\mu$m to keep the number of antidots reasonably small. At 4.2\,K the resonator perforated with three rows of antidots had a $Q$ factor of 10$^4$ at a perpendicular field of 1\,mT, decreasing to 5000 in a field of 4\,mT. The zero field $Q$ was $1.5 \cdot 10^4$. Orienting the field almost parallel to the substrate plane leads to $Q = 10^4$ at a field of 80\,mT and to $Q=5000$ at 110\,mT.  These numbers are likely to improve strongly when going to Millikelvin temperatures and using nm sized antidots.

\section{Perspectives and Conclusion}

In the present article, we have proposed a new hybrid system, namely a single electron above superconducting structures, which will allow to look deeply inside the physics of superconducting materials at high magnetic fields and the basic interaction between a single quantum optical system and a system with many degrees of freedom. The range of parameters to sense magnetic fields generated by the superconducting structures is certainly within the reach of modern experimental techniques of trapping and quantum manipulation of single electrons. It is also possible to attain the strong coupling regime between a single electron and the electric field of a single photon inside a superconducting resonator. That opens the possibility to explore a single trapped electron as a quantum memory for superconducting devices as well as to consider an array of individually trapped electrons as a quantum processor.

\emph{Acknowledgements}. FSK acknowledges financial support by the German Israel science foundation. PB, DK, RK and FSK acknowledge support by the German science foundation within the transregio TRR21. PB also acknowledges support of EU project MIDAS. DB acknowledges financial support by the Evangelisches Studienwerk Villigst e.V.. JN and MK acknowledge support by the Carl-Zeiss-Stiftung. AL, KI and MS acknowledge partly support by the DFG Center of Functional Nanostructures. We thank H. H\"affner for reading the manuscript.


%

\end{document}